\documentclass[prx,twocolumn,amsmath,amssymb,tightenlines,epsfig,superscriptaddress,nofootinbib,showkeys]{revtex4-2}

\usepackage{graphicx}
\usepackage{color}
\usepackage[percent]{overpic}
\usepackage{circledsteps}

\newcommand{\ket}[1]{|\,{#1}\,\rangle}
\newcommand{\braket}[2]{\mbox{$\langle\,{#1}\, | \,{#2}\,\rangle$}}

\newcommand{\sub}[2]{{#1}_{\mbox{\!\! \scriptsize #2}}}

\newcommand{\bv}[1]{\mathbf{ #1 }}

\def\beq{\begin{equation}}
\def\eeq{\end{equation}}

\def\CR{\nonumber\\[0.15cm]}

\newcommand{\rref}[1]{Ref.~\cite{#1}}
\newcommand{\fref}[1]{Fig.~\ref{#1}}
\newcommand{\frefp}[2]{Fig.~\ref{#1}~(#2)}

\newcommand{\eref}[1]{Eq.~(\ref{#1})}

\newcommand{\cref}[1]{chapter~\ref{#1}}

\newcommand{\Cref}[1]{Chapter~\ref{#1}}

\newcommand{\aref}[1]{appendix~\ref{#1}}
\newcommand{\bref}[1]{(\ref{#1})}

\usepackage{ulem}
\begin{document}

\title{In situ observation of chemistry in Rydberg molecules within a Bose-Einstein-condensate}
\author{Felix Engel}
\thanks{These authors contributed equally to this work.}
\affiliation{Physikalisches Institut and Center for Integrated Quantum Science and Technology, Universit{\"a}t Stuttgart, Pfaffenwaldring 57, 70569 Stuttgart, Germany}
\author{Shiva Kant Tiwari} 
\thanks{These authors contributed equally to this work.}
\affiliation{Department of Physics, Indian Institute of Science Education and Research, Bhopal, 462066, India}
\author{Tilman Pfau}
\affiliation{Physikalisches Institut and Center for Integrated Quantum Science and Technology, Universit{\"a}t Stuttgart, Pfaffenwaldring 57, 70569 Stuttgart, Germany}
\author{Sebastian~W\"uster}
\email{sebastian@iiserb.ac.in}
\affiliation{Department of Physics, Indian Institute of Science Education and Research, Bhopal, 462066, India}
\author{Florian Meinert}
\email{f.meinert@physik.uni-stuttgart.de}
\affiliation{Physikalisches Institut and Center for Integrated Quantum Science and Technology, Universit{\"a}t Stuttgart, Pfaffenwaldring 57, 70569 Stuttgart, Germany}

\begin{abstract}
We often infer the state of systems in nature indirectly, for example, in high-energy physics by
the interaction of particles with an ambient medium.
We adapt this principle to energies $9$ orders of magnitude smaller, to classify the final state of exotic molecules
after internal conversion of their electronic state, through their interaction with an ambient quantum fluid, a~Bose-Einstein condensate (BEC). 
The BEC is the ground-state of a million bosonic atoms near zero temperature, and a single embedded ultra-long range Rydberg molecule can coherently excite waves in this fluid, 
which carry telltale signatures of its dynamics. Bond lengths exceeding a micrometer allow us to observe the molecular fingerprint on the BEC \textit{in situ}, via optical microscopy.
Interpreting images in comparison with simulations strongly suggests that the molecular electronic state rapidly converts from the initially excited \textit{S} and \textit{D} orbitals to a much more complex molecular state (called ``trilobite''), marked by a maximally localized electron. This internal conversion liberates energy, such that one expects final-state particles to move rapidly through the medium, which is however ruled out by comparing experiment and simulations. The molecule thus must strongly decelerate in the medium, for which we propose a plausible mechanism.
Our experiment demonstrates a medium that facilitates and records an electronic state change of embedded exotic molecules in ultra-cold chemistry, with sufficient sensitivity to constrain velocities of final-state particles.
\end{abstract}

\maketitle

\section{ Introduction}
%
We excite extreme states of atoms and molecules, with a highly excited Rydberg electron orbiting far from the parent ion, within an extreme state of matter, a dilute gas Bose-Einstein condensate (BEC) \cite{Cornell_Wieman_BEC_RevModPhys} near the absolute zero of temperature. This already led to new insights into cold quantum chemistry \cite{schlagmueller:ucoldchemreact:prx}, electron atom scattering \cite{schlagmuller2016probing}, molecular physics \cite{zuber_molbondionryd}, and polaron formation \cite{camargo_polarons_exp_PRL}. It also challenges our understanding 
of the joint quantum dynamics of a few crucial actors, the Rydberg electron, ion and bound ground state atoms, with thousands of condensed
atoms in the BEC that surrounds these. Besides providing molecular constituents at very low energies, the latter was proposed as interrogation medium, akin to interrogation instruments used in high-energy physics. Similarly, the BEC can record cold molecular quantum dynamics and Rydberg chemistry \cite{mukherjee:phaseimp,Karpiuk_imaging_NJP,Tiwari_tracking}.

Here we report an initial experimental realization of this platform, using a BEC to herald the internal conversion of an electronic bound state, initialized in a Rydberg atomic orbital, into a Rydberg molecular trilobite orbital \cite{greene:ultralongrangemol,eiles2019trilobites,Shaffer_ucoldrydmolreview}. These exotic molecules are bound through elastic collisions of a highly excited Rydberg electron \cite{book:gallagher} with condensate atoms inside its orbital volume, and exist in a diverse zoo of molecular states \cite{greene:ultralongrangemol,eiles2019trilobites,Shaffer_ucoldrydmolreview}. The conversion is driven by nonadiabatic motion of the atoms in the condensate within the orbital sphere of the Rydberg electron on a microsecond timescale. It becomes evident when comparing the \textit{in situ} condensate density evolution, optically recorded through high-resolution phase contrast imaging \cite{Phase_Contrast}, with an extensive simulation campaign. These simulations are self-consistently predicting the motion of condensate atoms triggering conversion and can rule out survival of the initially excited electronic state, its conversion to other candidate molecules (butterflies) and motion of final state molecules. The latter requires rapid deceleration of molecules in the ambient quantum fluid, for which we propose a possible mechanism.

The experiment establishes BEC as probe medium for dynamical processes in ultracold atomic physics and chemistry, ranging from polaron formation  \cite{grusdt2017strong,camargo2018creation,schmidt2018theory}, angular momentum changing collisions \cite{Tiwari_tracking,schlagmueller:ucoldchemreact:prx}, the dynamical formation of ultra-long range molecules
\cite{greene:ultralongrangemol,schlagmueller:ucoldchemreact:prx,Luukko_polytrilob_PRL} or localized states \cite{Matthew:Localize}, over ionization \cite{Li_Gallagher_Ion_PRL,Amthor_mech_ion,Matthieu_Melting_PRA,park:dipdipionization} to phonon-mediated Yukawa interactions \cite{wang_rydelecBEC_PRL}. It also demands theoretical advances in several directions, to more rigorously investigate the mechanisms of condensate driven internal molecular conversion and trilobite molecule slow-down that we propose here.

\begin{figure*}[h!tb]
\includegraphics[width=2\columnwidth]{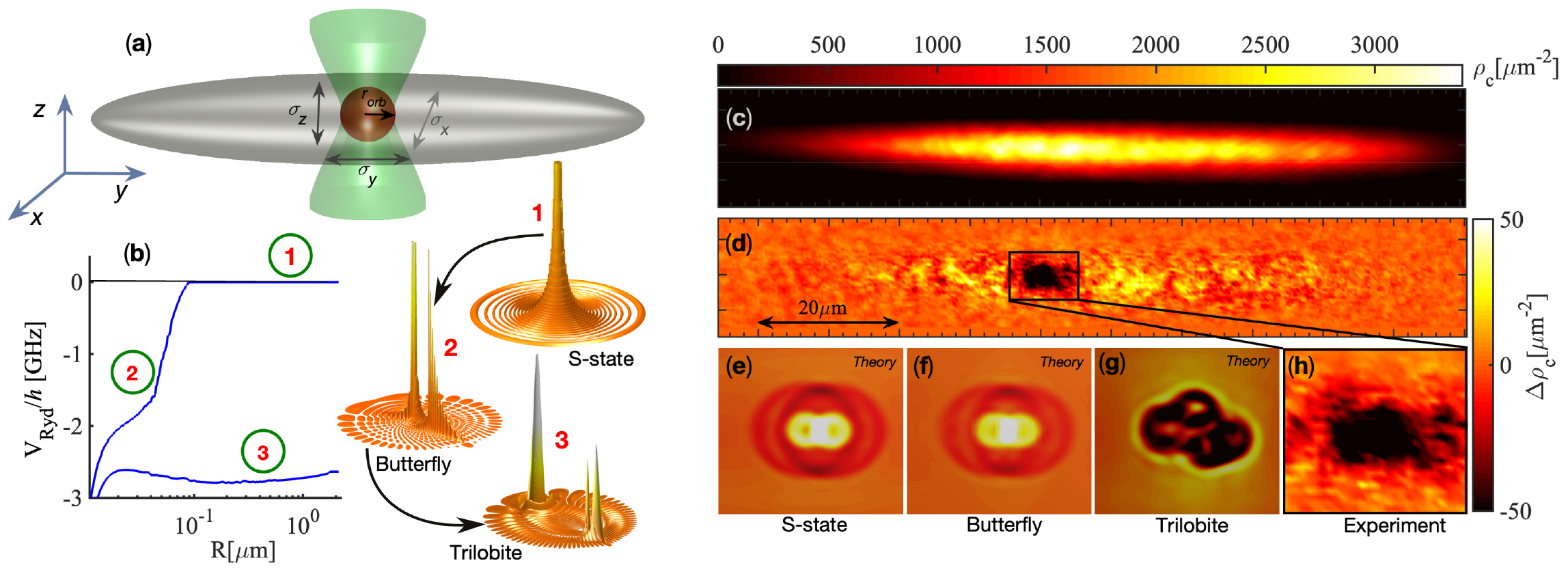}
\caption{\label{sketch} {\bf Rydberg excitation and subsequent electron dynamics in a  BEC.} (a) Single Rydberg atoms are created in the center region (red) of a cigar-shaped BEC (gray ellipsoid) using a focused laser beam (green), targeting $\ket{133S}$ and $\ket{130D}$ Rydberg states, with quantization axis along $y$. (b) Relevant molecular potential energy curves for the Rydberg electron interacting with the condensate, enabling state conversion from $S$ \Circled{1} (or $D$) Rydberg orbitals via butterfly molecular states \Circled{2} into trilobite molecular states \Circled{3}, with corresponding electron densities shown in orange.  (c) Phase contrast image of the bare column density $\rho_{\rm{c}}$ (density integrated along $z$) after interaction of a $133S$ Rydberg electron with the BEC for $\sub{N}{Ryd}=4$ repetitions at time $\sub{t}{evol}=200 \, \mu$s. (d) Data from (c) but with the BEC background in the absence of Rydberg excitations ($\rho_{\rm{bg}}$) subtracted, yielding the change of the column density $\Delta \rho_{\rm{c}} = \rho_{\rm{c}} - \rho_{\rm{bg}}$ due to the impact of the Rydberg electron inside the marked region. (e-g) Numerical simulations of the expected averaged signature $\Delta \rho_{\rm{c}}$ in the ambient BEC medium when the Rydberg electron remains in the initial $S$ state (e), transforms into a butterfly state (f) or transforms into a trilobite state (g). Each scenario causes a different imprint on the BEC, and only conversion into the trilobite state (g) explains the averaged experimental observation (h), a depression in the density instead of an increment. All simulations include a spatial uncertainty for the initial excitation of the Rydberg impurity, with standard deviations $\sigma_{x,y,z}=(0.9,2.6,0.9)$ $\mu$m; see panel (a).}
\end{figure*}

\section{ Rydberg electron imprint on a BEC} 
%
Our experiments start with a cigar-shaped BEC of typically one million $^{87}$Rb atoms in the hyperfine level $|F=2, m_F = 2\rangle$. It is confined in a magnetic trap with frequencies $\omega_{x,y,z} = 2\pi \times (194,16,194)\text{ Hz}$, resulting in a typical peak density of $\rho_0 = 4.6\times 10^{2} \, \mu$m$^{-3}$. Near the trap center, we promote a single atom from the BEC to a high-$n$ ($n$ is the principal quantum number) Rydberg state via resonant two-photon laser coupling to either $\ket{S}\equiv\ket{n=133,S_{1/2}, m_j=1/2}$ or $\ket{D}\equiv\ket{n=130,D_{5/2}, m_j=1/2}$ orbitals, see \aref{experimental_app}.
One of the excitation laser beams is strongly focused into the condensate via a high numerical aperture asphere (NA=0.55), constraining the spot of Rydberg atom creation to be as small as the size of the electron orbit (\frefp{sketch}{a}). After the $\sub{t}{exc} = 1 \, \mu$s long laser pulse, we let the electron interact for $\sub{t}{int} = 3 \, \mu$s with the $\approx 10^4$ condensate atoms residing inside its giant orbital ($\sub{r}{orb}=1.8 \, \mu$m radius for $n=133$). During this time, the electron may predominantly stay in its initial state or, as we discuss below, may undergo dramatic state changes due to its interaction with the ambient BEC (\frefp{sketch}{b}). The interaction is interrupted by applying a short ionizing electric field pulse ($7$ V/cm applied for $0.5 \, \mu$s), which rapidly removes the electron and ion fragments from the BEC. A short wait time of $0.3 \, \mu$s ensures that the electric field drops back to zero. This sequence of Rydberg creation, interaction, and removal from the BEC may be repeated $N_{\rm{Ryd}}$ times with a $208$ kHz repetition rate to increase the impact of the Rydberg electron on the BEC density. The latter is finally imaged (within a duration of $10 \, \mu$s) via a high-resolution \textit{in situ} phase contrast method through the same lens as used for local excitation \cite{Phase_Contrast}. An additional variable delay time of free condensate evolution is added between the end of the last Rydberg removal and the beginning of image acquisition to strengthen the signal by giving condensate atoms affected by interaction with the Rydberg atom more time to move. We denote the total evolution time from the start of the first excitation pulse to the beginning of imaging with $\sub{t}{evol}$.

The impact of the Rydberg electron on the BEC is barely visible on absolute scales (\frefp{sketch}{c}) and requires careful subtraction of the unperturbed background density. To this end, we take a reference image of the condensate for each experimental run using the same sequence but in the absence of Rydberg excitations, by tuning the excitation lasers $+100$ MHz off-resonant from the Rydberg state energy. Subtracting these reference images and subsequent averaging of typically 70-200 experiments reveals the impact of the Rydberg electron in the change of the condensate column density $\Delta \rho_{\rm{c}}$ (\frefp{sketch}{d}). This procedure also ensures correct subtraction of small density perturbations induced by the focused excitation laser. The experiment shows a local density reduction ("hole") near the Rydberg excitation region, which we now interpret by comparing to numerical simulations.

\begin{figure}[htb]
\includegraphics[width=\columnwidth]{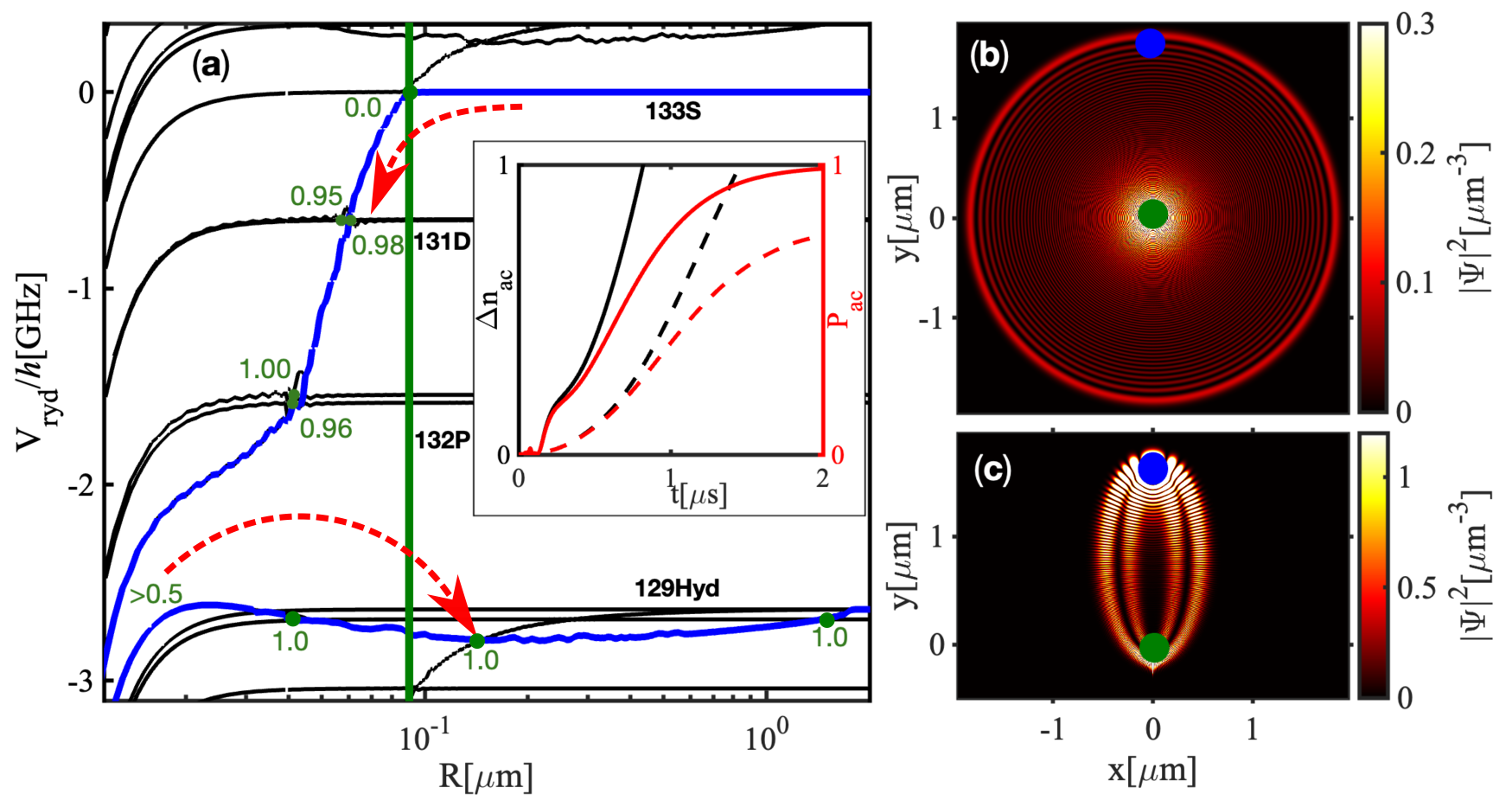}
\caption{\label{potentials}  {\bf Interactions of Rydberg electron and condensate atoms} (a) Molecular potential energy curves (black and blue lines) for a single ground state atom interacting with the Rydberg electron as a function of distance $R$ from the ionic core. Zero energy is that of $133S$ at large $R$. The ground state atom is expected to follow the blue curves along the red arrows once it reaches the avoided crossing (vertical green line at $R_{\rm{ac,s}}$), very likely transforming the electronic state from the initial $133S$ state to the $129$-trilobite state via butterfly states, based on Landau-Zener diabatic crossing probabilities shown in green. The inset shows the mean increase in the number of atoms $\Delta n_{\rm{ac}}$ (black) within the volume $R<R_{\rm{ac}}$ and the corresponding probability $P_{\rm{ac}}$ that at least one atom has entered this volume (red) as predicted from mean field simulations of the BEC following excitation to $\ket{S}$ (solid) or $\ket{D}$ (dashed). (b,c) 2D cut at $z=0$ through the 3D electron probability density for the $133S$ molecular state (b) with attached condensate atom at $R=1.68 \, \mu$m, and the $129$-trilobite state (c) at $R=1.58 \, \mu$m. The green (blue) circle marks the position of the Rydberg core (attached condensate atom).} 
\end{figure}
%

\section{ BEC dynamics and electron state change} 
%
Once the high-$n$ Rydberg state is excited in the BEC, about 10.000 condensate atoms will reside in the orbital volume of the Rydberg electron. Each of these atoms then interacts with the electron via low-energy $\rm{s}$- and $\rm{p}$-wave collisions \cite{Fermi:Pseudo,greene:ultralongrangemol}. For an atom at position $\bv{R}$ relative to the Rydberg ionic core, that collision is quantified by respective scattering lengths (and volumes for p) $a_{\rm{s},\rm{p}}[k(\bv{|R|})]$, which depend on the local electron momentum $\hbar k$ and thus implicitly on the distance $R=|\bv{R}|$ from the ion core \cite{Omont:Pwave} (see \aref{potentials_app}). When assuming that the electron remains in the initially excited $\ket{S}$ ($\ket{D}$) orbital, each condensate atom experiences a single molecular potential energy surface $\sub{V}{ryd}(\mathbf{R})$ for a large part of its configuration space, which can be derived perturbatively using $\rm{s}$-wave collisions alone, yielding $\sub{V}{ryd}^{(S,D)}(\mathbf{R})= V_0(\bv{R})|\psi_{S,D}(\bv{R})|^2$, with $\psi_{S,D}(\bv{R})=\braket{\bv{R}}{S,D}$ the electron wavefunction of the respective Rydberg orbital and $V_0(\bv{R}) = 2\pi\hbar^2 a_{\rm{s}}[k(\bv{|R|})]/m_e$ ($m_e$ is the electron mass). Such a potential is net attractive for both $\ket{S}$ and $\ket{D}$, since $a_{\rm{s}}<0$ for most Rb-e$^-$ collisions relevant here. This results in an overall increase of BEC density at the center (\frefp{sketch}{e}) \cite{Tiwari_atomsinatoms_NJP}, just the opposite to our experimental observation (\frefp{sketch}{h}).

This simplified picture with a single potential surface changes, when taking into account $\rm{p}$-wave collisions. The $\rm{p}$-wave scattering channel features a shape resonance, which leads to a divergence of $a_{\rm{p}}$. The resonant $\rm{p}$-wave interaction can then strongly mix the highly degenerate Rydberg orbitals with  angular momentum $l>2$, forming butterfly molecular states \cite{Hamilton_2002:Butterfly}. Those exhibit strong avoided crossings with the potential surfaces $\sub{V}{ryd}^{(s,d)}$ associated with the initially excited $\ket{S}$ ($\ket{D}$) orbitals at $R_{\rm{ac,s}} \approx 0.1 \, \mu$m  ($1900 a_0$) [$R_{\rm{ac,d}} \approx 0.05 \, \mu$m  ($945 a_0$)], opening channels for rapid internal conversion of the electronic state of the Rydberg molecule and ultracold chemistry following laser excitation \cite{schlagmueller:ucoldchemreact:prx} (\textit{cf.}~\frefp{sketch}{b}). Lastly, the butterfly states couple at short range to trilobite molecular states \cite{schlagmueller:ucoldchemreact:prx}, which feature strongest localization of the Rydberg electron on a single condensate atom \cite{greene:ultralongrangemol}, on multiple atoms \cite{Eiles_highdensenv_2016}, or even clusters of atoms \cite{Luukko_polytrilob_PRL}. The corresponding potential energy landscape for one condensate atom interacting with the Rydberg electron is obtained via diagonalizing the electron-atom interaction Hamiltonian on a finite Rydberg atomic basis (see \aref{potentials_app}), and shown in \frefp{potentials}{a} together with arrows that indicate the discussed path for Rydberg state changes. Approximate diabatic Landau-Zener crossing probabilities are indicated in green in \fref{potentials}, for the butterfly to trilobite crossing extrapolating \cite{schlagmueller:ucoldchemreact:prx}, and support the conversion sequence discussed. Experimental evidence for this has been found in Ref.~\cite{Geppert_diffusive_redist_NatComm}.

\begin{figure*}[h!tb]
\includegraphics[width=2\columnwidth]{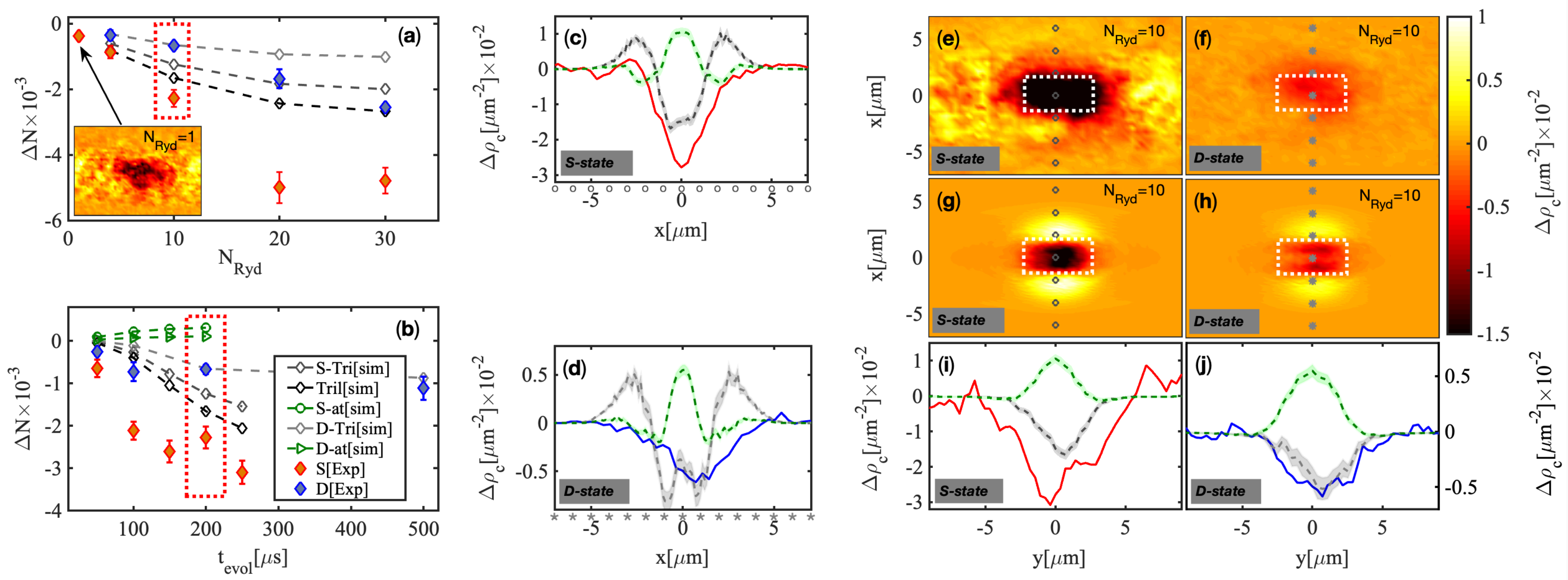}
\caption{\label{hole_creation} {\bf Dynamical evolution of the averaged BEC density induced by Rydberg excitations to $133S$ and $130D$.} (a,b) Measured (full symbols) and simulated (open symbols connected by lines) depression of the BEC density quantified by the local atom number change $\Delta N$, of the order of a few thousand, in the Rydberg excitation region (white boxes in (e-h)). $\Delta N$ is shown as a function of the number of sequentially implanted Rydberg atoms $N_{\rm{Ryd}}$ with $t_{\rm{evol}} = 200 \, \mu$s (a), and as a function of the time since first implantation  $t_{\rm{evol}}$ for $N_{\rm{Ryd}}=10$ (b). The inset in panel (a) shows the effect of a single Rydberg excitation, the legend in panel (b) also pertains to panel (a). Simulation data depict scenarios for which the electron remains in the initially excited $S$-state ($D$-state) (green in (b)), converts from $133S$ to $129$-trilobite after $\sub{t}{conv,s}=0.7$ $\mu$s (dashed dark gray with $\diamond$ marked $S$-Tri), converts from 130D to 128-trilobite states after $\sub{t}{conv,d}=2$ $\mu$s (dashed light gray with $\diamond$ marked $D$-Tri), or converts into a trilobite without delay $\sub{t}{conv,s}\approx0$ (dashed black with $\diamond$ marked Tril). The red-dotted boxes surround data for which details are shown exemplarily in panels (c)-(j). (e)-(h) Measured (e),(f) and simulated (g),(h) change of the column density $\Delta \rho_{\rm{c}}$ following $S$- and $D$-excitations as indicated, for $\sub{N}{Ryd}=10$ and $t_{\rm{evol}} = 200 \, \mu$s. Simulation data are shown for delayed trilobite insertion (at $\sub{t}{conv,s/d}$), and account for finite optical resolution. (c),(d), (i),(j) Vertical ($y=0$) (c,d) and horizontal ($x=0$) (i,j) cuts through the measured (solid lines) and simulated (dashed lines) column density differences $\Delta \rho_{\rm{c}}$, with markers below panels (c,d) indicating the cut in panels (e-h). (c-i) follow $S$-excitation, and (d,j) $D$-excitation. (c),(d),(i),(j) Green dashed lines assume electrons staying in the initial electronic states, and gray dashed lines assume delayed conversion into trilobite molecules. 
Experimental data are typically averaged over $70-200$ runs, while the simulations are averaged over $200$ samples, the standard error of which is indicated by colored shading in panels (c),(d),(i),(j). Error bars in panels (a),(b) show one standard error.
}
\end{figure*}

In a next step, we analyze the BEC dynamics to assess the possibility that these additional potential energy curves may become relevant. First, we simulate the time evolution using the Gross-Pitaevskii-equation (GPE) for the mean-field BEC wavefunction $\phi$ with an external potential given by the asymptotic $\ket{S}$ or $\ket{D}$ potentials in \frefp{potentials}{a}. Incorporating the experimental sequence of multiple Rydberg excitation pulses yields the predicted increase of the BEC density, $\rho=|\phi|^2$, at the Rydberg excitation region mentioned above [\textit{cf.}~\frefp{sketch}{e}], contradicting our experiment. Yet, this simulation now provides an estimate for the timescale of Rydberg electron state change into the butterfly and then trilobite potential. Specifically, we evaluate the time $t=\sub{t}{conv}$, when the probability $P_{\rm{ac}}(t)$ approaches unity, that at least one atom has passed the avoided crossing with the butterfly state from outside to inside. We mark $R_{\rm{ac,}s}$ as green line in \frefp{potentials}{a} and 
the inset of \frefp{potentials}{a} shows $P_{\rm{ac}}(t)$. It is calculated from the net change of mean atom number within $R<\sub{R}{ac}$, $\Delta n_{\rm{ac}}(t) = \int_{R<R_{\rm{ac}}} d^3 \mathbf{R} [\rho(\bv{R},t)-\rho(\bv{R},0)]$, see \aref{GPE_app}. At this moment, which we set to $\sub{t}{conv,s}=0.7$ $\mu$s after excitation for $\ket{S}$, and slightly later $\sub{t}{conv,d}=2.0$ $\mu$s for $\ket{D}$, rapid state change of the Rydberg electron first into the butterfly and then into the trilobite state occurs. Once in the trilobite, the Rydberg electron strongly localizes at the position of a specific condensate atom [\frefp{potentials}{c}], which causes a much more potent but shorter-range potential seen by the BEC.

The discussion so far considered only one atom in the electron orbit. While interaction potential calculations for all 10.000 condensate atoms in the orbit remain intractable, we provide evidence in \aref{manybody_app} and \aref{electron_stripoff_app} for the following picture: (i) initially, if the Rydberg atom was excited, all BEC atoms are located away from the shell near $R=R_{\rm{ac}}$, since the molecular states in that shell have strongly reduced $\ket{S}$ character, (ii) outer condensate atoms then slowly move inwards due to net attraction, correctly modelled in the GPE. (iii) As soon as \emph{the first} condensate atom reaches $R\approx R_{\rm{ac}}$, it may adiabatically take the Rydberg electron to the trilobite molecular state. (iv) Despite each molecular constituent gaining an initial velocity of $v=3.5$ m/s in the process, kinetic energy is rapidly shed by interactions with the many-body environment, in which the electron repeatedly re-attaches to new, previously immobile, ground-state atoms.
 We thus arrive at a phenomenological model for GPE simulations, in which the condensate atoms only initially experience the potential due to an electron in an $\ket{S}$ ($\ket{D}$) atomic Rydberg state. After a conversion time $\sub{t}{conv,s}$ ($\sub{t}{conv,d}$), we switch to the potential associated with the relevant \emph{immobile} trilobite molecular potential with bond length $1.58$ $\mu$m and random orientation. We thus neglect short lived transient states during conversion and
assume a Landau-Zener probability of $100\%$ for trilobite formation for simplicity, since accurate formation probabilities are challenging to calculate in the high density medium (see \aref{manybody_app}). The result of such a simulation for $\ket{S}$ [\frefp{sketch}{g}] predicts a clear density reduction, in qualitative and near quantitative agreement with the experiment.
For this, large forces had to be exerted on BEC atoms for a prolonged time, necessitating an immobile trilobite electron.
In contrast, even trilobite motion as slow as $v=0.5$ m/s reverts the signal back to a density increment. 

The response of the condensate medium involves outwards propagating matter waves that interfere with each other and also represent condensate particles interfering with themselves. Both features cannot be captured by a classical treatment, see supplementary movie. While our experimental resolution was not yet high enough to read out the information in interference effects, this could be realized in the future. Properties of the BEC that have been already exploited here were the absence of thermal density fluctuations which helped the extraction of the very weak signal created by the Rydberg impurities, and the relatively large densities which triggered the molecular conversion.

\section{Quantitative analysis of the BEC response} 
%
Next, we test our phenomenological model through a quantitative comparison with experimental data taken after $\ket{S}$ and $\ket{D}$ Rydberg excitation at different values for $N_{\rm{Ryd}}$ and $\sub{t}{evol}$. In all cases, we observe a local density reduction, which we quantify further by evaluating the corresponding atom number change $\Delta N$ in the excitation region [\frefp{hole_creation}{a) and (b}]. It is evident that the impact of the Rydberg electron on the BEC increases monotonically in $N_{\rm{Ryd}}$ and $\sub{t}{evol}$ similarly for $\ket{S}$ and $\ket{D}$. We also show that the data does not agree with simulations in which the Rydberg electron state does not change, supporting the conclusion that both states undergo molecular state changes into similar trilobite states, akin to internal molecular conversion. The quantitatively weaker condensate response to $\ket{D}$ excitations is due to slower conversion $(\sub{t}{conv,d}>\sub{t}{conv,s})$, which we attribute to the nonradial gradient of the $\ket{D}$-state potential (see \aref{potentials_app}). 
Noteworthy, even the faint impact of just a single Rydberg atom ($N_{\rm{Ryd}}=1$ in \frefp{hole_creation}{a}) is still detectable. 

A comparison of exemplary experimental \textit{in situ} images for $\ket{S}$ and $\ket{D}$ excitations with simulations using \eref{impurityGPE} to model the phenomenological sequence proposed above is shown in \fref{hole_creation}(c)-(j). Both the width and amplitude of the hole like feature appearing in $\Delta \rho_{\rm{c}}$ is captured well. This implies that atoms have been strongly accelerated by interaction with the Rydberg electron, ultimately leaving the excitation region. We stress again, that this acceleration cannot come from an electron remaining in the initially excited atomic state, which then should show a local density increase instead of a hole. Similarly, we can exclude butterfly molecules as final state (\textit{cf.}~\frefp{sketch}{f}). We have already ruled out several other conceivable causes for the qualitative mismatch between the experiment and simulations retaining the electronic initial state, such as three-body losses, non-mean field dynamics and short range physics of the interaction potential in \rref{Tiwari_atomsinatoms_NJP}. To conclude, our observations require the deep potential associated with the extreme localization of the Rydberg electron wavefunction in trilobite states, which can sufficiently strongly accelerate condensate atoms in the excitation region to create a ``hole'' feature. The trilobite molecules must quickly become immobile to explain the data, since motion as slow as $v=0.5$ m/s turns the hole feature in simulations into a density increase, see \aref{imaging_app}.

We provide a candidate mechanism for trilobite slowdown in \aref{electron_stripoff_app}: By re-attachment of the Rydberg electron onto new condensate atoms, after the first attached condensate atom has been propelled out of the orbital volume, the ion interacts with a large number of ambient condensate atoms, itself being slowed down to a complete stop in less than half a microsecond.

The data in \fref{hole_creation} shows that the mechanism proposed above results in near quantitative agreement with the experiment.
Many other conceivable scenarios are ruled out by our simulations. That retention of the originally excited electronic states
is not possible is explicitly shown in the figure. Moving Rydberg atoms in S,D or other adjacent low-l states yield similar qualitative disagreement. Earlier experiments in \rref{schlagmueller:ucoldchemreact:prx} rule out  transitions to lower principal quantum numbers than $n-4$.

We justify our use of trilobite wavefunctions derived for a single atom interacting with the Rydberg electron by the observation of \rref{Luukko_polytrilob_PRL} that trilobite orbitals in a high density medium typically are of similar character as single atom ones.
 However it is likely that many-body effects quantitatively modify the mechanism. For example the trilobite electron could be localized on multiple condensate atoms; see, e.g., Ref.~\cite{Eiles_highdensenv_2016}. As a first check of such possibilities, we obtain similar simulation results from the trilobites shown here as from a coherent superposition of two trilobites, with attached condensate atoms an angle $\theta=\pi/6$ apart.

While the atom number change $\Delta N$ of the order of thousands is quantitatively explained by our self consistent model following $\ket{D}$ excitation, the agreement is less good following $\ket{S}$ excitation. It is somewhat improved by assuming immediate conversion to the trilobite. Possible trilobite creation via $S$-excitation in dense gases has been proposed in Ref.~ \cite{Luukko_polytrilob_PRL}, through admixture of $S$ character into trilobites by atom clusters in dense media. Extrapolating calculations of Ref.~\cite{Luukko_polytrilob_PRL} (that neglect the p-wave shape resonance), this effect should occur for $n=133$  only at higher densities $\rho= 1.13\times 10^{21}$m$^{-3}$. However, since the shape resonance increases the mixing of states, considering clusters formed near $R\approx R_{\rm{ac}}$ could conceivably lower the required densities. Exploring this poses a formidable challenge for theory, since the dense condensate atoms in 3D at $n=133$ necessitate diagonalizing \eref{S_P_wave} in the complete Hilbert space, with $M=17689$ bare atomic states in the $n=133$ manifold alone.  

Another aspect in which the experiment severely challenges theory, is the quantum dynamics of thousands of condensate atoms all jointly coupled to the same Rydberg electron and thus entering correlated many-body states. While we were still able to describe observations using existing techniques, the model was not based on first principles and is not likely to remain successful for refined experiments with more detailed observation: Even if advanced bases \cite{Eiles_highdensenv_2016} ultimately yield complete 3D many-body potentials, each avoided crossing passed in \frefp{potentials}{a}, will split the Rydberg electron wavefunction onto multiple surfaces, subsequently entangling with the BEC, necessitating advanced multi-orbital methods 
\cite{mlmctdhb_kronke2013,mlmctdhb_jcp, mlmctdhb_ion_impurity,pendse_PhysRevA}. These superpositions then decohere \cite{rammohan:tayloring,rammohan:superpos}, which in turn can affect transitions at subsequent crossings. The details will affect final observations, for example through orientation of trilobite molecules. 

Considering the above, the level of agreement between simulation and experiment is encouraging. 
Seeking even more complete agreement in the future will allow the ultracold medium tracking platform to provide constraints on future theories
designed to tackle those challenges.

\section{ Outlook and conclusions}
%
We report imaging the imprint of a molecular state conversion on the quantum dynamics of an ambient Bose-Einstein condensate, \textit{in situ}, with phase contrast imaging.
By comparison with simulations for alternative scenarios of molecular dynamics, we show that the data constrains timescales, velocities and final quantum states involved in the conversion process, demonstrating the utility of the BEC as an interrogation medium, similar in functionality to high energy physics detectors or reaction microscopes. We had shown in Ref.~\cite{Tiwari_tracking} that coherence also has the potential to allow the BEC to record nonzero velocity.
Here it was used already to infer an unexpectedly vanishing velocity.
Our conclusions heavily rely on extensive and detailed simulations, in much the same way that the interpretation of collision data in high energy physics requires accurate Monte-Carlo event generators \cite{Mangano_eventMC_review}, which allow the prediction of observations based on known physics and from that the isolation of signals for new physics.
Here, in much the same spirit the ultra-cold ``interrogation device'' has already guided us to propose a new mechanism for the slowdown and preservation of trilobite molecules in a BEC.

With further improvements in the spatio-temporal resolution of density measurements, the platform may allow direct access to ultracold quantum chemical reactions of exotic Rydberg molecules and their interplay with a quantum degenerate solvent, transient aspects of polaron formation \cite{grusdt2017strong,camargo2018creation,schmidt2018theory}, many-body trilobite molecules \cite{Luukko_polytrilob_PRL}, phonon mediated Yukawa interactions \cite{wang_rydelecBEC_PRL} or Rydberg composites \cite{Hunter_Composites_PhysRevX}.
To unleash the full potential of the condensate interrogation medium for few body quantum dynamics prototyped here then will require also the further development of theory techniques that can handle the simultaneous interaction of many ground state atoms with one Rydberg electron and their entanglement with the latter.

\acknowledgments
The authors thank S.~Dutta, M.~Eiles, J.~M.~Rost, R.~Schmidt and M.~Wagner for fruitful discussions. We acknowledge support from Deutsche Forschungsgemeinschaft [Projects No. PF 381/13-1 and No. PF 381/17-1, the latter being part of the SPP 1929 (GiRyd)], and financial support as well as computational resources at the MPCDF from the Max-Planck society under the MPG-India partner group program.  F. M. is indebted to the Baden-W\"{u}rttemberg-Stiftung for the financial support by the Eliteprogramm for Postdocs. The project received funding from the European Research Council (ERC) under the European Union's Horizon 2020 research and innovation programme (Grant Agreement No.~101019739 - Lon-gRangeFermi).

\appendix
%
\section{ Rydberg excitation in the BEC}
\label{experimental_app}
%
The two-photon excitation exploits the intermediate $6P_{3/2}$ level, using lasers at about 420 nm and 1011 nm. The intermediate state detuning is set to $+160$ MHz ($+320$ MHz) for the $133S_{1/2}$ ($130D_{5/2}$) data. 
Typical BEC atom numbers are $N=1.2 \times 10^6$ for $S$-state data and $N=0.7 \times 10^6$ for $D$-state data, resulting in peak densities $\rho_0 = 4.9 \times 10^2 \mu\text{m}^{-3}$ and $4.0 \times 10^2 \mu\text{m}^{-3}$, respectively.
The 1011 nm beam propagates along the $z-$direction and is focused into the BEC center. The 420 nm beam illuminates the BEC homogeneously and propagates along the $y$-direction (quantization axis set by the trap magnetic field) for addressing the $S$-state, while for the $D$-state the beam is counter-propagating to the 1011 nm beam. The laser polarization of the 420 nm and 1011 nm beam is set to $\sigma^+$ $-$ $\sigma^-$ for the $S$-state and to $\pi$ $-$ $\pi$ for the $D$-state data, which allows for coupling selectively to the $m_j = 1/2$ Rydberg state in both cases. The tight 1011 nm beam waist of $w_0=1.8(3) \, \mu$m enables localization control of the implanted Rydberg atom. The interaction of the Rydberg electron with the condensate atoms inside the electron orbit during the laser excitation causes density-dependent line shifts and broadening. To compensate for that, we detune the 1011nm laser by $-50$ MHz for $133S_{1/2}$ and $-30$ MHz for $130D_{5/2}$ from the bare Rydberg state resonance, which approximately accounts for measured mean density shifts. The strong Rydberg blockade and low excitation rate ensure that only a single condensate atom is excited to a Rydberg state for each pulse \cite{schlagmuller2016probing}.

\begin{figure}[htb]
\centering
\includegraphics[width=1.0\columnwidth]{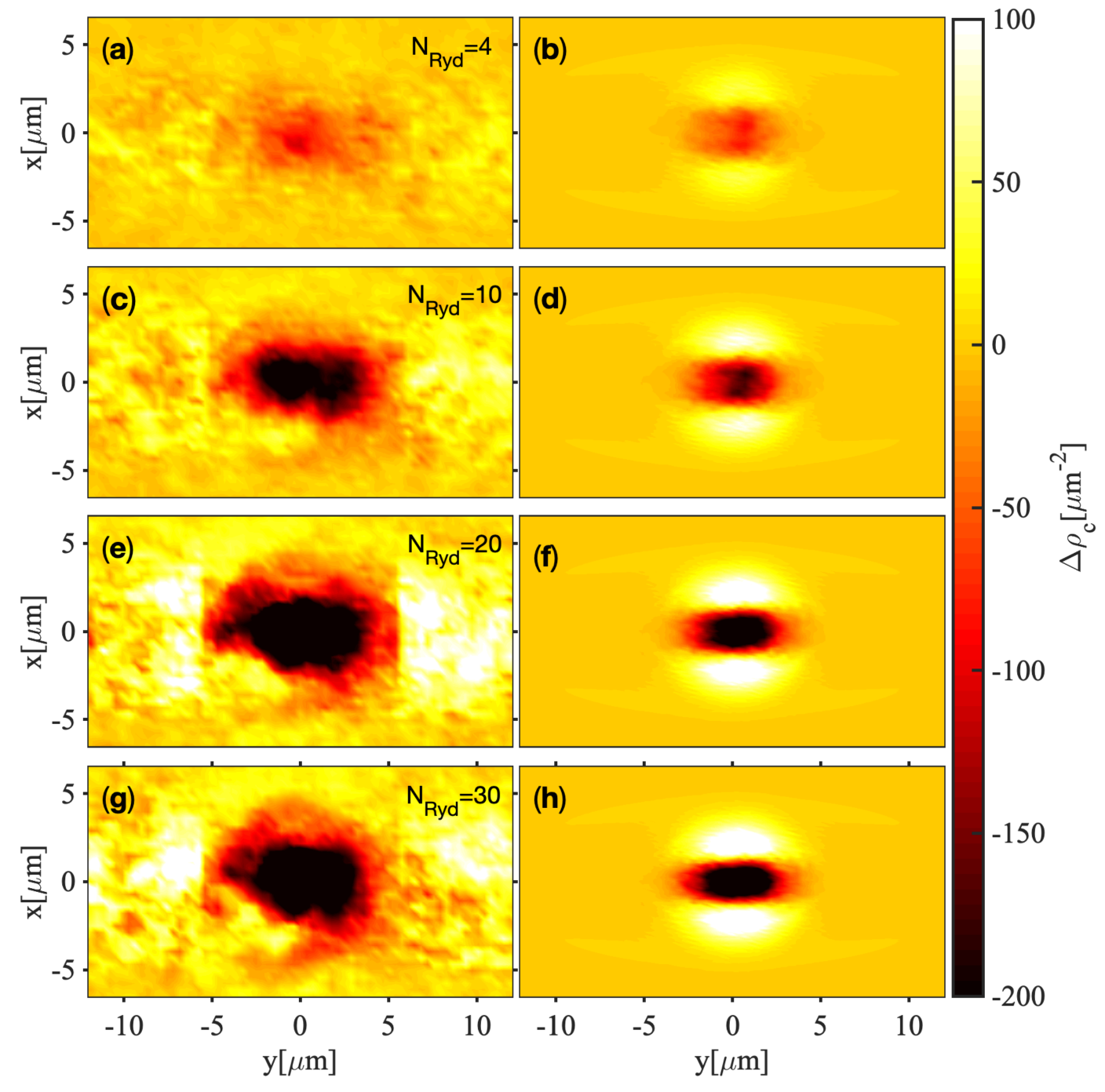}
\caption{\label{SI_sstate_pulsenumbers} {\bf Averaged density depression following repeated $\ket{S}$ Rydberg excitations in a BEC}.
(a),(c),(e),(g) The relative column density in the experiment for $\sub{N}{Ryd}$ repeated excitations as indicated, at $\sub{t}{evol} = 200\mu$s, and (b),(d),(f),(h) row wise matching simulation results for the same, assuming conversion to the trilobite molecule from the initial $\ket{S}$ state after $0.7$ $\mu$s.}
\end{figure}
\begin{figure}[htb]
\centering
\includegraphics[width=1.0\columnwidth]{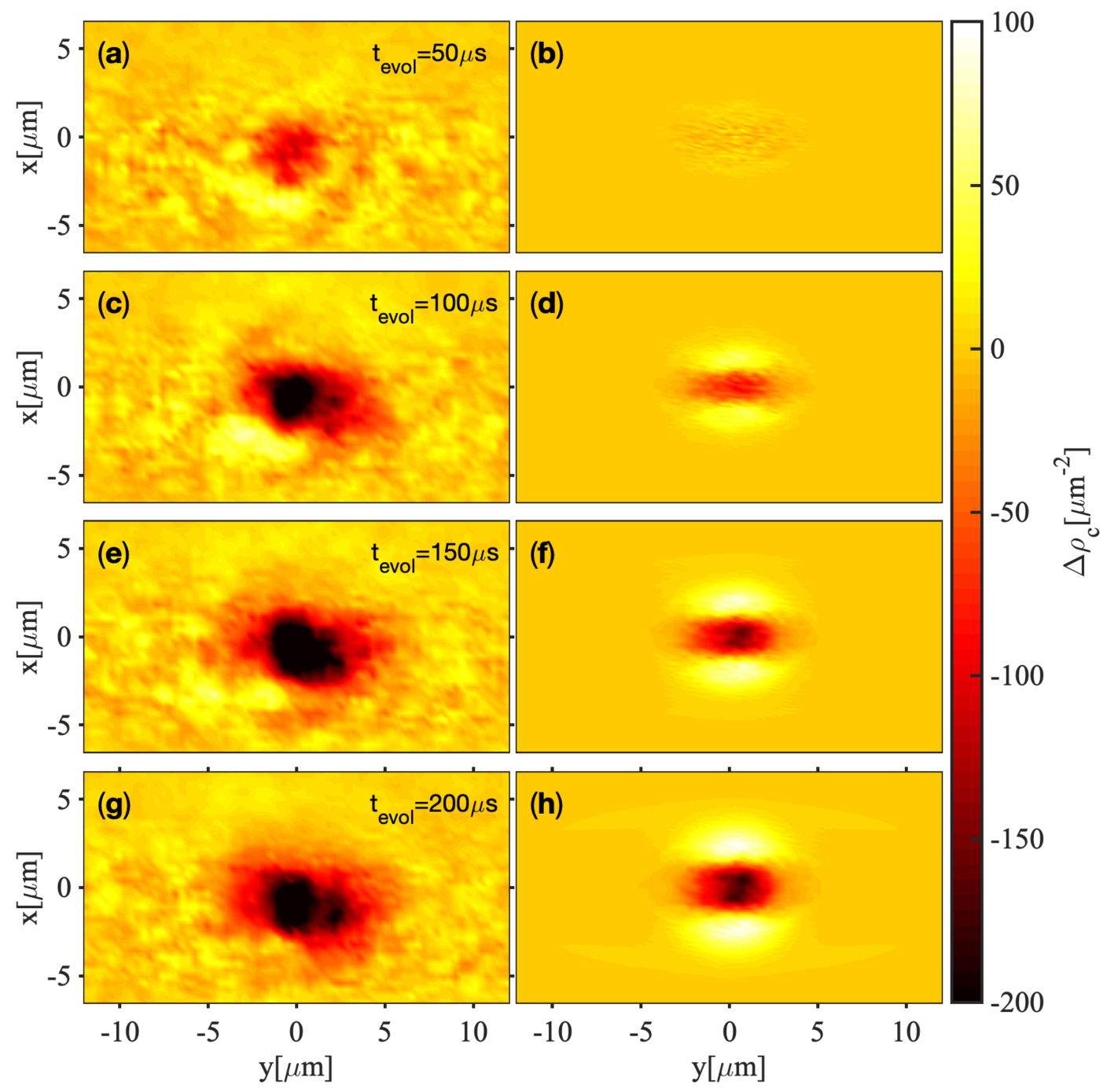}
\caption{\label{SI_sstate_evoltimes} {\bf Averaged density depression following repeated $\ket{S}$ Rydberg excitations in a BEC}. The same as \fref{SI_sstate_pulsenumbers}, but at fixed $\sub{N}{Ryd}=10$ for different wait times.}
\end{figure}
%
\section{\textit{In-situ} imaging and data processing}
\label{imaging_app}
%
For phase contrast imaging, we illuminate the BEC with a $780$ nm imaging beam propagating along the $z$-direction and detuned by $+500$MHz from the $\ket{5P_{3/2},F=3,m_F=2}$ state. For this detuning and our typical densities, the BEC induced phase shift on the imaging light is in the linear response regime \cite{Phase_Contrast}, which allows us to directly infer the integrated column density $\rho_c$ and from that $\Delta\rho_c$. Small shot-to-shot position and atom number fluctuations of the BEC make it difficult to use a single reference image for this procedure. Instead, we construct an optimal background reference for each image of $\rho_c$ taken with Rydberg excitations present. This reference is obtained from a large pool of recorded background density distributions (typically 200 images) following the procedure described in Ref.~\cite{Ockeloen_smallnumberdetection_PhysRevA}. When constructing the reference, we exclude a region of $10 \times 10 \, \mu$m around the Rydberg excitation region. Experimental data shown are typically averaged over $70-200$ runs and $280$ runs in \fref{sketch}.

\begin{figure}[htb]
\centering
\includegraphics[width=1.0\columnwidth]{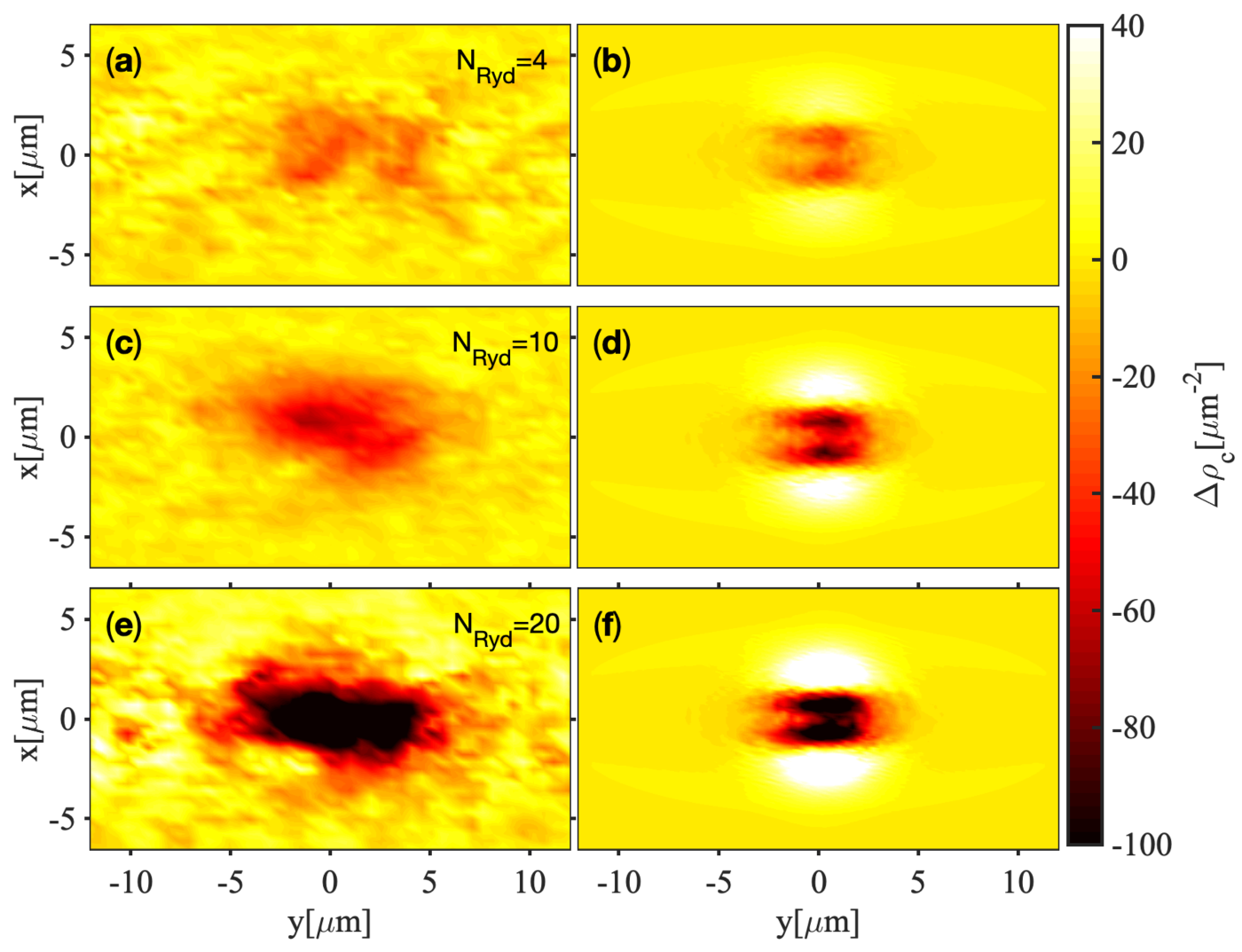}
\caption{\label{SI_dstate_pulsenumbers}{\bf Averaged density depression following repeated $\ket{D}$ Rydberg excitations in a BEC}.
(a,c,e) The relative column density in the experiment for $\sub{N}{Ryd}$ repeated excitations as indicated, at $\sub{t}{evol} = 200\mu$s, and (b,d,f) row wise matching simulation results for the same, assuming conversion to the trilobite molecule from the initial $\ket{D}$ state after $2$ $\mu$s. }
\end{figure}

\begin{figure}[htb]
\centering
\includegraphics[width=1.0\columnwidth]{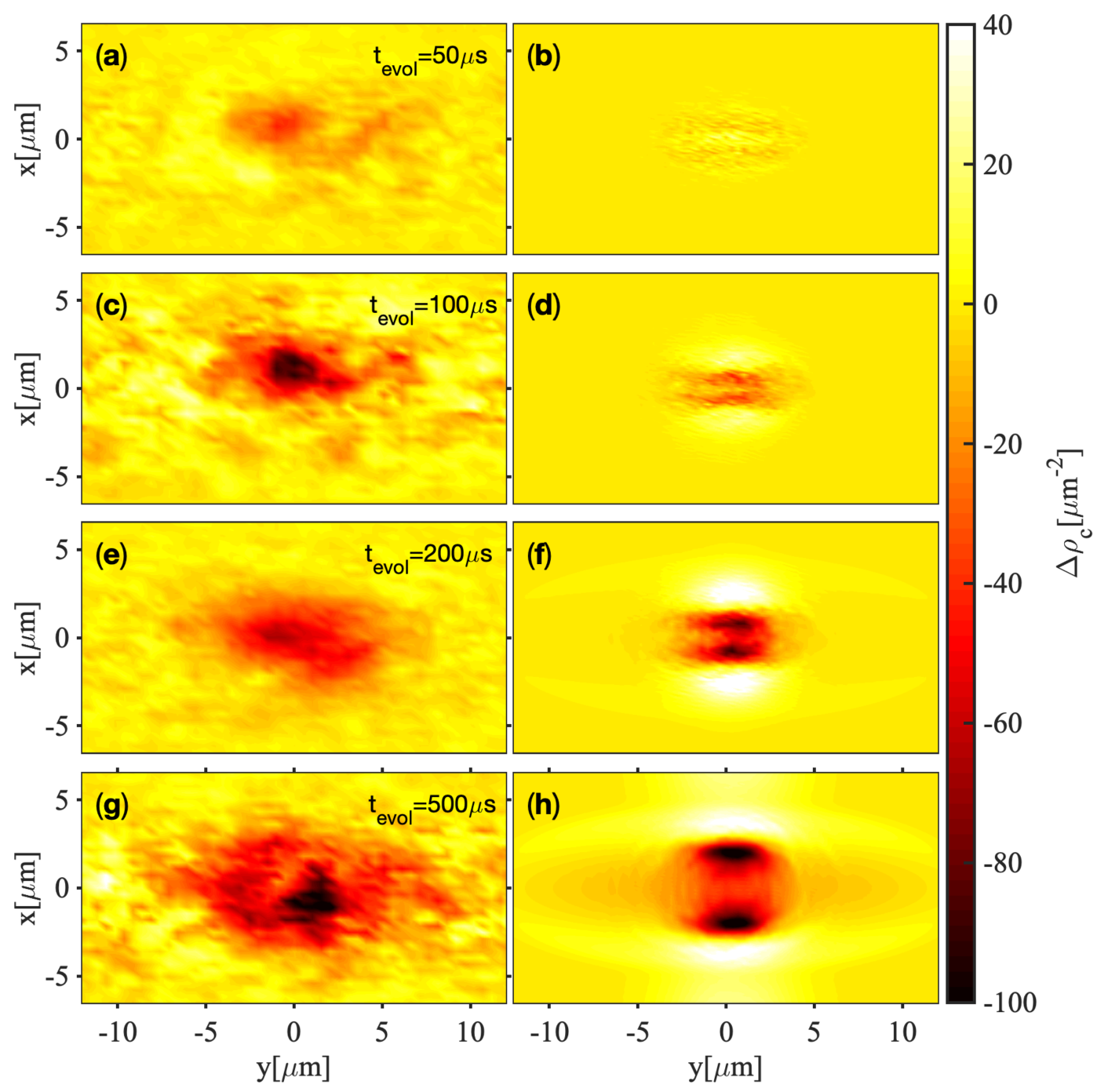}
\caption{\label{SI_dstate_evoltimes} {\bf Averaged density depression following repeated $\ket{D}$ Rydberg excitations in a BEC}. The same as \fref{SI_dstate_pulsenumbers}, but at fixed $\sub{N}{Ryd}=10$ for different wait times.}
\end{figure}

Further snapshots of condensate density evolution beyond \fref{hole_creation} can be found in \fref{SI_sstate_pulsenumbers} - \fref{SI_dstate_evoltimes} and exhibit a similar level of agreement with simulations. They all show the general trend that the relative density depression increases in size and amplitude with more repeated Rydberg excitations $\sub{N}{Ryd}$, or longer final free evolution time $\sub{t}{wait}=\sub{t}{evol}-\sub{N}{Ryd}\sub{t}{pulse}$, with $\sub{t}{pulse}=4.8$ $\mu$s the total duration of the sequence for a single Rydberg excitation. Some of the experimental samples correspond to a near vanishing wait time, and demonstrate that also directly after the excitation the signal corresponds to a hole like feature, which simply slightly widens during the final wait period.

\begin{figure}[htb]
\centering
\includegraphics[width=1.0\columnwidth]{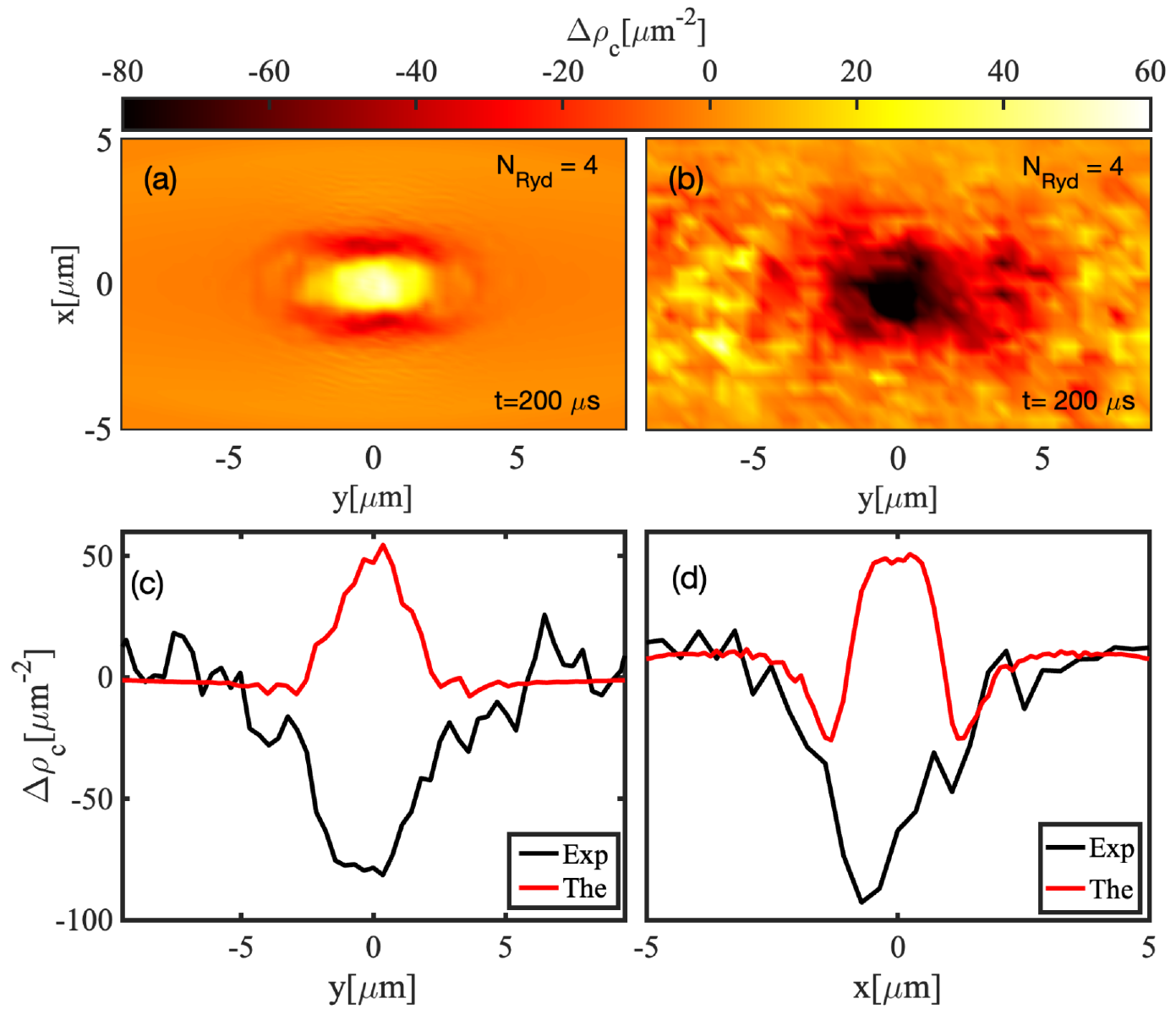}
\caption{\label{SI_motion} {\bf Expected signature of moving trilobite molecules}. 
Averaged density depression, after excitation of $\ket{S}$ states for  $\sub{N}{Ryd}=4$ and $\sub{t}{evol}=200$ $\mu$s. (a) Relative column density change $\Delta \rho_{\rm{c}}$ from simulation with trilobites inserted at $t=0$ with random orientation and moving with $v=0.5$ m/s in the direction of their internuclear axis, does not agree with the experiment in panel (b). (c) Horizontal ($x=0$) and (d) vertical ($y=0$) cuts through the measured (black) and simulated (red) column density changes $\Delta \rho_{\rm{c}}$.}
\end{figure}

Finally we show a comparison of experiment and simulation in \fref{SI_motion} that rules out moving trilobites: We can see that already motion as slow as $v=0.5$ m/s (instead of the $3.5$ m/s expected after $S$ state to trilobite conversion) causes the BEC response to change from the density depression (black line) to a density increase in the center (red line), which is in qualitative disagreement with experiments. The reason is that a mobile trilobite affects the condensate wavefunction over a larger region than an immobile one but much more weakly, such that condensate atoms are not sufficiently accelerated to leave the imaging region. 
%

\section{Rydberg - condensate potentials}
\label{potentials_app}
%
Electron-atom interactions are described by pseudo potentials \cite{Fermi:Pseudo,greene:ultralongrangemol}:
 \begin{align}
 \label{FullPot}
 &\mathcal{V}(\bv{R}_n,\bv{r}) = \frac{2 \pi \hbar^2}{m_e} \bigg( a_s[k(|\bv{R}_n|)] \delta^{(3)}(\bv{R}_n-\bv{r})\CR
 & + 3a_p[k(|\bv{R}_n|)]\delta^{(3)}(\bv{R}_n-\bv{r})  \overleftarrow{\boldsymbol{\nabla}}_{\bv{r}}\cdot \overrightarrow{\boldsymbol{\nabla}}_{\bv{r}} \bigg),
 \end{align}
 where $a_{s,p}[k]$ are triplet s- or p-wave scattering lengths \cite{Omont:Pwave} for electron - $^{87}$Rb collisions. The electron momentum $\hbar k$ is taken from $ \hbar^2 k(\bv{R}_n)^2/(2m_e) = E_{\nu^*} + e^2/(4\pi \epsilon_0 |\bv{R}_n|)$, where $E_{\nu^*} = -\sub{R}{Ryd}/n^{*2}$ is the electron energy for effective principal quantum number $n^*$ \cite{Gallagher:Qdefect1, Gallagher:Qdefect2}, $m_e$ is the electron mass and $\sub{R}{Ryd}$ the Rydberg constant. Scattering phases are taken from Ref.~\cite{Bahrim_2001}. 

We then add the ion-atom potential \cite{Fontana:Ion1, Flannery_2005:Ion2} and $\hat{V}=\sum_n  \mathcal{V}(\bv{R}_n,\bv{r})$ from \eref{FullPot} to the unperturbed $^{87}$Rb atomic Hamiltonian, and diagonalize the resultant $\sub{\hat{H}}{el}(\bv{R})$ for each collective coordinate $\mathbf{R}=[\mathbf{R}_1,\cdots,\mathbf{R}_N ]^T$ of  condensate atoms:
\begin{eqnarray}
\sub{\hat{H}}{el}(\bv{R})\ket{\varphi(\bv{R})}&=\sub{V}{ryd}(\bv{R})\ket{\varphi(\bv{R})},
\label{S_P_wave}
\end{eqnarray}
in the set of atomic states with $n=127-135$ and all angular momenta, to find a large number of Born-Oppenheimer surfaces $\sub{V}{ryd}(\bv{R})$.
A single attached atom can reside on the quantization axis, such that only states containing $m_\ell=0,\pm1$ are required. Since we only excite $m_j=1/2$, and $m_j$ remains a good quantum number when neglecting condensate spin flips, we further restrict these to $m_j=1/2$.
This is not possible for multiple condensate atoms in 3D, rendering this scenario too challenging to calculate potentials for $n=133$. For simulations we thus take either (i) the single atom surface connecting to $\ket{S}$ at large $|\mathbf{R}|$, (ii) the approximate single atom surface connecting to $\ket{D}$ at large $|\mathbf{R}|$ discussed below, 
(iii) the surface corresponding to dimer trilobite molecules $\sub{V}{ryd}(\bv{R})= \frac{2 \pi \hbar^2}{m_e}  a_s[k(|\bv{R}|)]  |\sub{\psi}{Tril}(\bv{R})|^2$, with wavefunction $\sub{\psi}{Tril}$ given by the analytical formula from \rref{Eiles_highdensenv_2016}:
\begin{eqnarray}
\sub{\psi}{Tril} &=  \frac{u'_{n0}(t_{-}) u_{n0}(t_{+}) - u_{n0}(t_{-}) u'_{n0}(t_{+})}{4\pi \Delta t}.
\label{trilobite_wavefunction}
\end{eqnarray}
Here $\Delta t = t_{+}-t_{-}$, with $t_{\pm} = (R + r \pm \sqrt{R^2 + r^2 - 2Rr\cos(\gamma)}$, $r=|\mathbf{r}|$, $R=|\mathbf{R}|$, and $\gamma$ the angle between the ground-state atom at $\mathbf{R}$ and Rydberg electron at $\mathbf{r}$. $u_{n0}(r)$ is the radial Hydrogen electron wavefunction in state $n$, $\ell=0$.
In a later section, we provide justifications for this approach by considering multiple condensate atoms on the $z$-axis.
While \eref{trilobite_wavefunction} is derived based on just a single perturbing atom, it has been numerically shown in Ref.~\cite{Luukko_polytrilob_PRL} that also in denser environment the typical electron distribution strongly resembles \bref{trilobite_wavefunction}, albeit now localized on dominant \emph{clusters} of atoms.

For (ii) we construct an approximation, since the computational requirement to attach atoms only on the $z$-axis prevents direct calculations of the anisotropic potential following excitation of $\ket{130D,J = 5/2,m_j = 1/2}$. Note that the state contains two different orbital angular momenta, 
$\ket{J = 5/2,m_j = 1/2}= \sqrt{\frac{3}{5}} \ket{ m_l = 0, m_s = 1/2} 
+ \sqrt{\frac{2}{5}} \ket{m_l = 1,m_s = -1/2}$.
But the electron density governing the interaction with the BEC differs in $\ket{130D, m_l = 0, m_s = 1/2} $ and $\ket{130D, m_l = 1,m_s = -1/2}$, decohering the superposition above, as discussed in \cite{rammohan:superpos,rammohan:tayloring}. We assume complete decoherence and average these two potentials, taking into account their different spin dependence. 

%
\begin{figure}[htb]
\centering
\includegraphics[width=1.0\columnwidth]{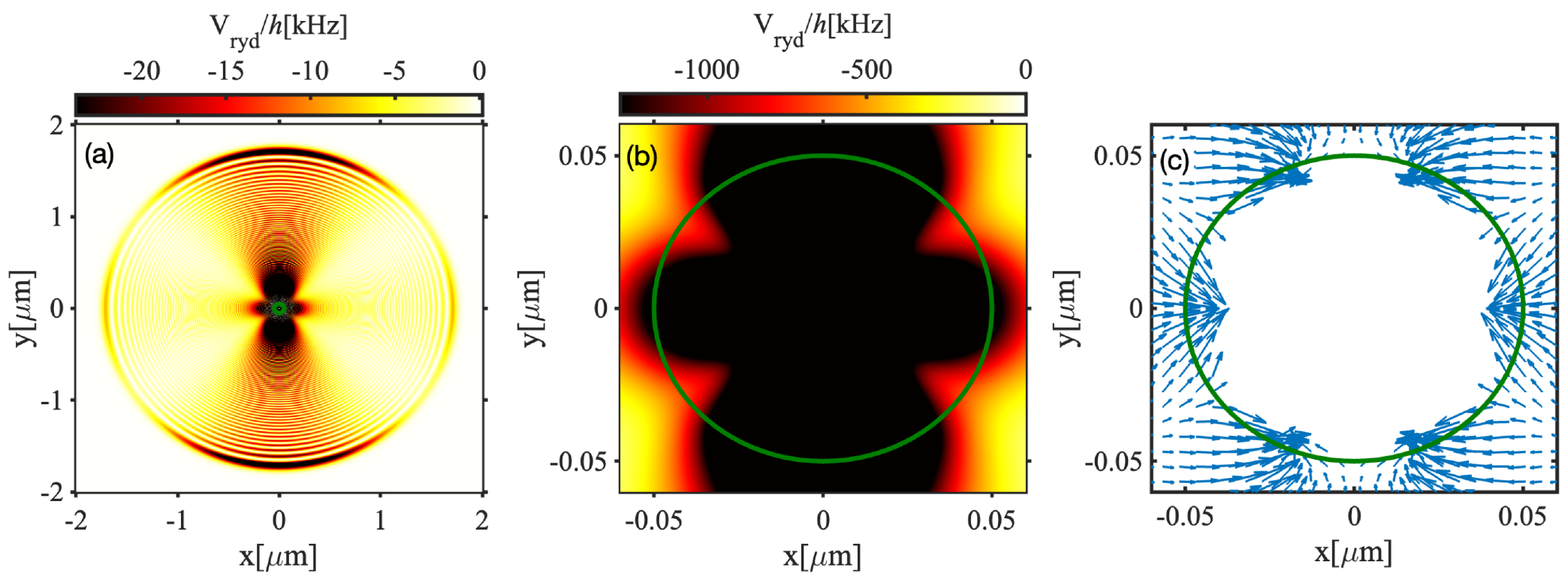}
\caption{\label{SI_dpotential_gradient} {\bf Approximate potential for Rydberg D state interacting with condensate}. (a) $\sub{V}{ryd}^{(d)}(\bv{R})$ in the x-y plane, where the quantisation axis is along y. (b) Zoom of the same near the avoided crossing, indicated by the green circle in (a,b). Here we apply Gaussian smoothing with kernel width $\sigma=0.05$ $\mu$m to focus on the net trend of the potential. (c) Gradient of the smoothed potential near the avoided crossing shown as arrows, to capture the net trend of $\boldsymbol{\nabla}\sub{V}{ryd}^{(d)}(\bv{R})$.
}
\end{figure}
For this, we first find $\sub{V}{ryd,T}^{[d(m_j=1/2)]}(z)$ for one atom on the $z$-axis asymptotic to $\ket{130D, m_J = +1/2}$ from \eref{S_P_wave}, using the triplet scattering length.
Since the $m_l = 1$ component vanishes on the $z$-axis, we extract the $m_l=0$ contribution through $\sub{V}{ryd,T}^{(d0)}=\frac{5}{3}\sub{V}{ryd,T}^{[d(m_j=1/2)]}$.
We then use this in 3D via $\sub{V}{ryd,T}^{(dm)}(\mathbf{R})=\sub{V}{ryd,T}^{(d0)}(R) |Y^{m}_{2}(\theta, \varphi)|^2/ |Y^{0}_{2}(0,0)|^2$, which assumes the radial potential to not differ between $m_l=0,1$ and the angular dependence to follow that of the $\ket{130D, m_l = m}$ electronic state. We also require the singlet contribution
$
\sub{V}{ryd,S}^{(d1)}(\mathbf{R}) \approx  \frac{a^{S}_s(0)}{a_s(0)}\sub{V}{ryd,T}^{(d1)}(\mathbf{R}),
$
using the zero energy singlet scattering length $a^{S}_s(0) = 0.63a_0$. 

We finally assemble
$
\sub{V}{ryd}^{(d)}(\bv{R}) = \frac{3}{5}\sub{V}{ryd,T}^{(d0)}(\mathbf{R})  + \frac{1}{5}\big(\sub{V}{ryd,S}^{(d1)}(\mathbf{R}) + \sub{V}{ryd,T}^{(d1)}(\mathbf{R})\big),
$
the sum in brackets taking into account that scattering between the  $m_s = -1/2$ Rydberg electron and $m_f=2 $ condensate atom will contain equal singlet and triplet parts. The resultant potential is shown in \fref{SI_dpotential_gradient}, including a zoom onto $\sub{R}{ac,d}$ to highlight the nonradial gradient.

\section{Many body Rydberg potentials} 
\label{manybody_app}
\begin{figure}[htb]
\includegraphics[width=1.0\columnwidth]{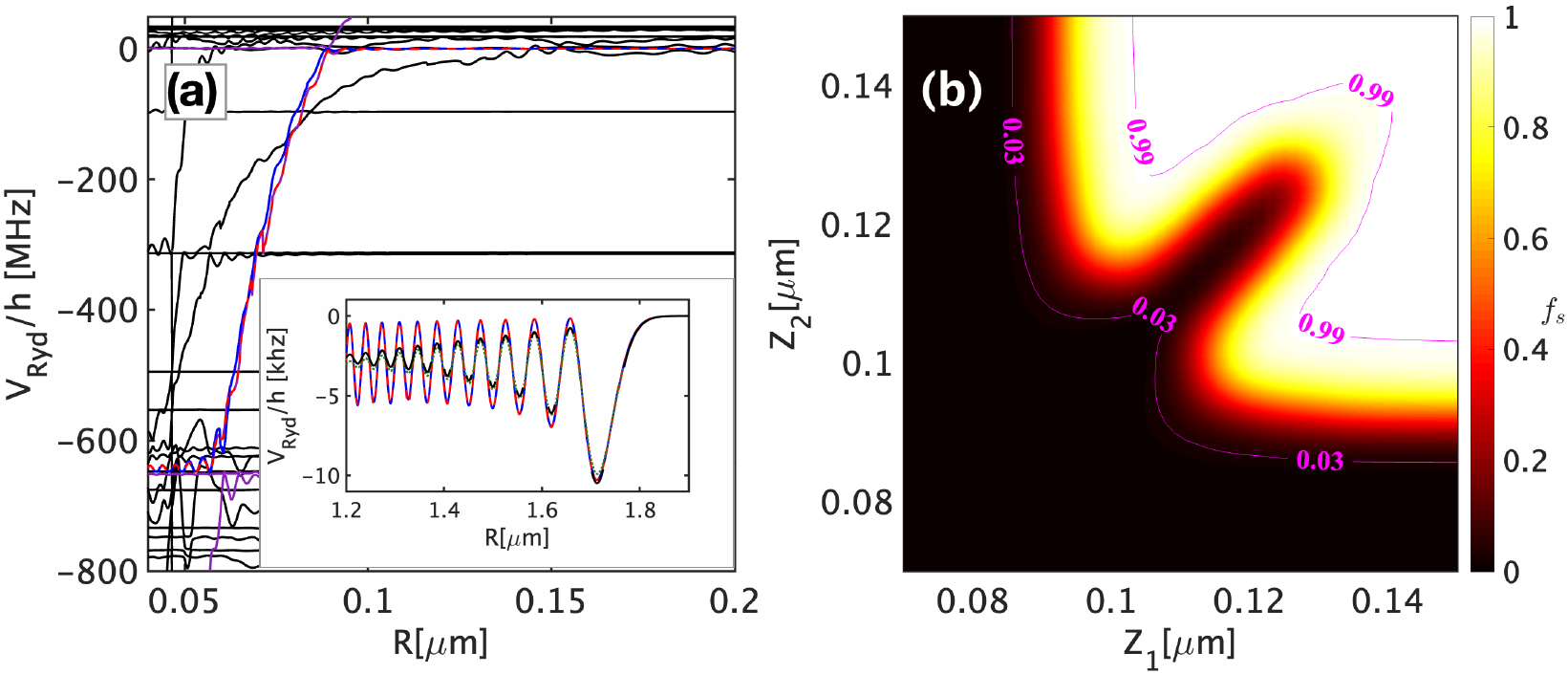}
\caption{{\bf Multiple condensate atoms interacting with the Rydberg electron}. \label{manybodypotentials} (a) Potential with $N_g=86$ additional atoms on the z-axis (blue and violet lines, with blue following the asymptotic $\ket{133S}$ state), compared to a single atom potential (red dashed lines), $1/3$ of the potential for a moving \emph{cluster} of three atoms (black lines), and in the inset the corresponding average of three single atom potentials (green dotted line). The inset shows the outer orbital region representing the by far largest volume fraction, where many and single body potentials nearly agree, and even moving clusters do not cause qualitative deviations. The main panel shows the inner volume near the shape resonance. Deviations between many body potential (blue) and single body potential (red dashed) are visible and could strongly alter Landau-Zener crossing rates. Deviations are even stronger for the moving cluster (black), which leads to an earlier onset of the butterfly crossing, but a shallower slope. (b) Fractional $s$-character $f_s(Z_1,Z_2)$ for two atoms on the $z$-axis. Between the magenta lines, the s-content of the initially excited state drops from $0.99$ to $0.03$. Again potentials are additive, unless both atoms approach $R_{ac}$ together.
 }
 \end{figure}

%
To justify our dominant use of potentials calculated for just a single atom interacting with the Rydberg electron, we now extend 
calculations to multiple interacting atoms. These are constrained to reside on the quantisation axis ($z$-axis), since otherwise calculations would not be tractable.
We distribute $N_g=86$ ground-state atoms randomly on the z-axis in configuration $\mathbf{R}_0=[\mathbf{R}_1, \cdots \mathbf{R}_{86}]^T$, which provides a similar mean nearest neighbor distance in 1D as would be the case for $\sub{N}{orb}\approx 10^4$ in 3D. We then scan one additional atom at $\mathbf{R}_s$ regularly along $z$ to map out the potential $\sub{V}{ryd}([\bv{R}_0,\mathbf{R}_s]^T)$. This yields the surface asymptotic to the $\ket{S}$ state shown in \fref{manybodypotentials}{a}. In that figure, we also compare this with a potential $\sub{V}{add}(\mathbf{R}_s)=\sub{V}{ryd}(\mathbf{R}_s)+\sub{V}{ryd}(\bv{R}_0)$, where the energy offset of the earlier $N_g$ atoms is just added to a single atom potential for the scanned atom. We find that both mostly agree, hence assuming additive potentials for all condensate atoms while the Rydberg electron is in low-l states, as implicitly done in \eref{impurityGPE}, is a good approximation for the overwhelming fraction of the 3D orbital volume. Deviations are only visible between many and single atom potentials near the avoided crossing $R=\sub{R}{ac,s}$, and become extreme when moving a cluster of three atoms. This will
affect the width of avoided crossings between molecular states, thus many-body effects can strongly modify the Landau-Zener conversion rates calculated in \cite{schlagmueller:ucoldchemreact:prx}. Since these are thus not accurately known, we assume $100\%$ conversion to the trilobite for simplicity.

We also scan the overall s-state character $f_s(Z_1,Z_2)=|\braket{133S}{\varphi(\bv{R})}|^2$, for two independent atoms located at $Z=Z_1$ and $Z=Z_2$, shown in \fref{manybodypotentials}{b}. The molecular state corresponds to $>99.5\%$ to the single atom $s$-state, unless either atom enters $Z<0.064$ $\mu$m. Only for \emph{both} atoms near $\sub{R}{ac}$, the additive picture fails.
Generalizing to the $3\sub{N}{orb}$ dimensional molecular energy surface for $\sub{N}{orb}$ atoms, we thus assume that when \emph{any of those enter} the sphere with radius $\sub{R}{ac}$ around the ion, it can trigger an adiabatic molecular transition into the trilobite, justifying our phenomenological model. The classical time to fall down the potential drop to the trilobite is less than $50$ ns once started, hence we neglect the transient butterfly state.

For high-l states or trilobite molecules potentials are not additive, however we can justify the use of two-body trilobites in our phenomenological model by the observation in \cite{Luukko_polytrilob_PRL}, that most many-body trilobites in a dense medium approximate two-body ones centered on clusters, see also \aref{electron_stripoff_app}.

\section{Condensate evolution and processing}
\label{GPE_app}
%
We insert the chosen $\sub{V}{ryd}$ as in (i-iii) above into the GPE
\begin{eqnarray}
\label{impurityGPE}
&i\hbar \frac{\partial}{\partial t}\phi= \bigg(-\frac{\hbar^2}{2 m}\boldsymbol{\nabla}^2 + W + U_0 |\phi|^2 +\sub{V}{ryd}  \bigg)\phi,
\end{eqnarray}
with 3D harmonic trap $W = m(\omega^2_x x^2 + \omega^2_y y^2 + \omega^2_z z^2)/2$. The strength of interactions is $U_0 = 4 \pi \hbar^2 a_b/m$, with $a_b = 109 a_0$ the s-wave scattering length of $^{87}$Rb atoms and $m$ their mass. We increase $\omega_y\rightarrow (2\pi)32$ Hz for tractability, keeping peak densities as in experiment.
Prior to the excitation sequence, the ground-state is found in imaginary time. For comparison with experiment, simulation data is post-processed assuming a finite imaging resolution of $1$ $\mu$m \cite{Tiwari_atomsinatoms_NJP}. We average all simulations over $200$ realisations of random Rydberg positions, which was only possible with $128^3$ grid-points. This will slightly undersample the most strongly accelerated atoms, possibly explaining discrepancies outside the white dashed box in \fref{potentials}.

From $\Delta \sub{n}{ac}$ defined in the main text, the probability $\sub{P}{ac}$ that at least one atom (out of $N$ total) has crossed the green line in \fref{potentials} from outside to inside is
$
\sub{P}{ac}(t)=\left[ 1- \left(1-\frac{ \Delta n_{ac}(t)}{N}\right)^{N} \right].
$
For the classical simulations in the supplementary movie, we solved Newton's equations for a large number of non-interacting background gas atoms initially randomly located within a cubical volume, and then subject to the same excitation sequence of Rydberg atoms and trilobites used for the GPE simulation shown in the movie. To calculate the classical forces, we use the numerical gradients of the relevant energy surface derived from \eref{S_P_wave}, and the analytical gradient of \eref{trilobite_wavefunction}, taken from \rref{Eiles_highdensenv_2016}.

\section{Electron strip-off and slowdown of Trilobite molecules}
\label{electron_stripoff_app}
%
The conversion from the initial electronic Rydberg state $\ket{133S}$ to the $\ket{129Hyd}$ trilobite molecule liberates $\Delta E/h\approx2.7$ GHz of potential energy into kinetic energy of ion and attached condensate atom. The $v_r=7$ m/s relative velocity should cause the condensate atom to leave the Rydberg orbit after $\sub{t}{exit}=0.27$ $\mu$s, whereupon one could expect the electronic state to turn into a set of hydrogenic states.

\begin{figure}[htb]
\includegraphics[width=1.0\columnwidth]{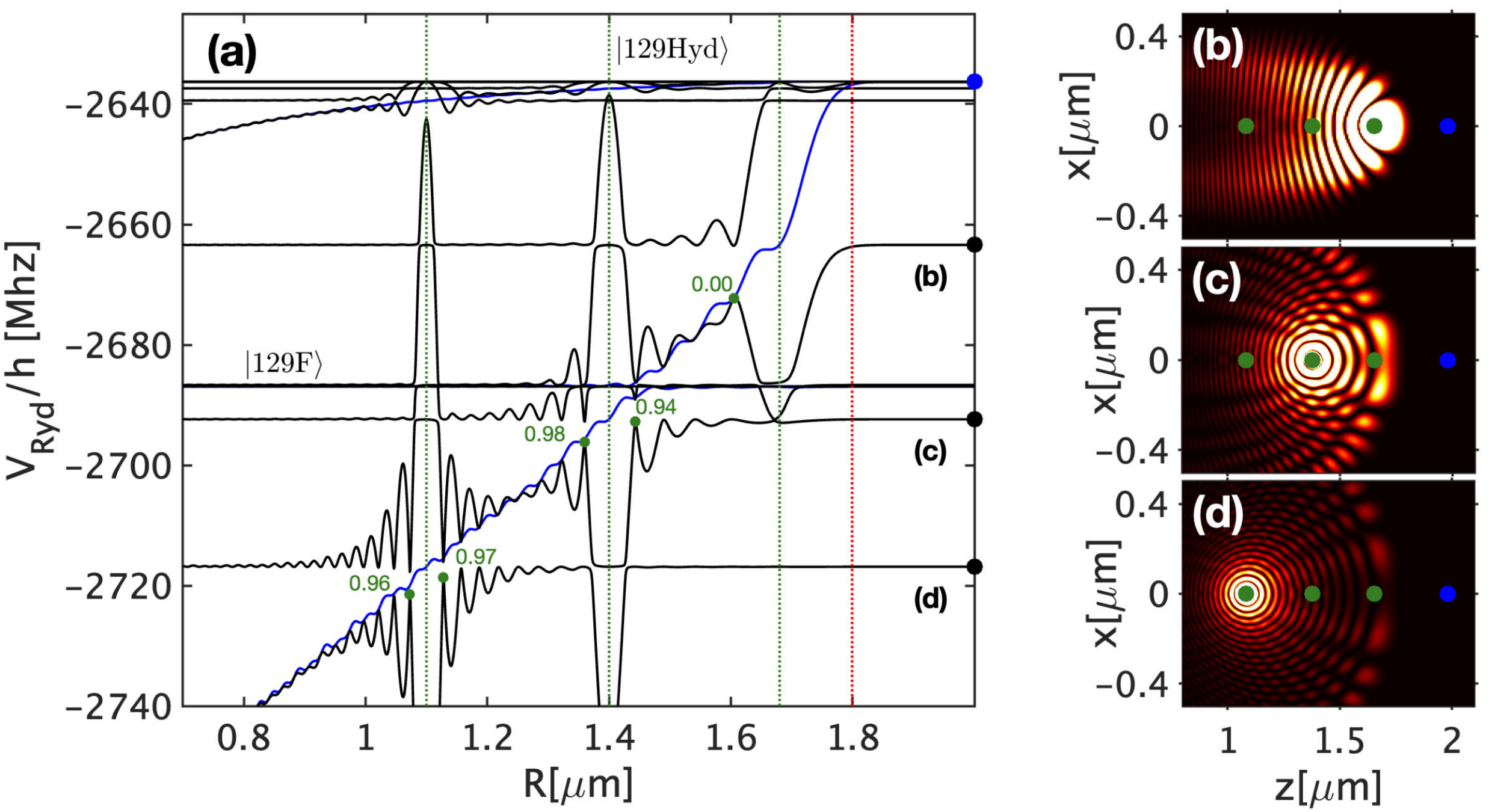}
\caption{{\bf Electron strip-off from moving attached atoms}. \label{trilobite_slowdown} (a) Rydberg potentials from \eref{S_P_wave}, zooming near $R\approx \sub{r}{orb}$ (vertical red dotted line) at the energy range in which the trilobite molecular potential connects to the hydrogenic manifold $\ket{129Hyd}$, in the presence (black) and absence (blue) of three extra perturbers fixed at $z=1.2,\:1.429,\:1.68$ $\mu$m, marked by green dotted vertical lines in (a), and green dots in (b-d). The position of a fourth perturber is scanned along $R$. The zero of energy is that of $\ket{133S}$. Green numbers are Landau Zener crossing probabilities as discussed in the text. (b-d) Electron density in the $xz$ plane, for the states marked in (a), with fourth atom at $R=2$ ${\mu}$m as shown by the blue dot.
 }
 \end{figure}

However this will not be the case in the BEC, as one sees from many-body potentials in the region where the trilobite potential would connect to the hydrogenic manifold, see \fref{trilobite_slowdown}. Without the BEC, the attached atom in the trilobite potential climbs a roughly linear potential gradient $\boldsymbol{\nabla} V\approx90$ MHz/$\mu$m following the blue potential line, until reaching the hydrogenic manifold. However when adding three representative static BEC atoms on the z-axis, each opens up a series of avoided crossings with the trilobite potential. For the relative velocity of $3.5$ m/s with which the accelerated perturber passes those initially at rest, most should be diabatically traversed as indicated by the estimates of Landau Zener crossing probabilities shown in green. The crucial exception is the last crossing, near the condensate atom situated most closely to the orbital exit. Here the electron is likely to ``strip off'' from the fast moving atom onto that last static atom, causing the loss of half the initial kinetic energy from the relative motion once the original perturber leaves the molecule. At the experimental density, the replacement atom is always available.

At this time, the ion is still moving. In the simplest picture, repeated relocations of the electron onto a new atom in the orbit cause the ion to continuously feel a constant force given by $\boldsymbol{\nabla} V$ above. However, in our dense medium, about 5 atoms are located within the region of strong trilobite potential, and scanning a cluster of five atoms steepens the gradient to $\boldsymbol{\nabla} V\approx600$ MHz/$\mu$m. With this force, a crude classical simulation of ion motion shows complete stopping within $1$ $\mu$m and $1$ $\mu$s. While this slowdown is not yet fast enough to explain observations, it provides a compelling candidate mechanism, quantitative details of which require a more complete 3D picture.
The ion could be additionally decelerated by elastic collisions with the surrounding ground-state atoms \cite{Dieterle_iontransp_PhysRevLett,Sourav_ioncooling_PhysRevA,Ravi_sadiq_ioncooling,Mahdian_swapcooling_NJP}.

Besides slowing down the ion, repeated electron relocation preserves the electronic state in a Trilobite for much longer than it would exist in vacuum (only $\sub{t}{exit}$), which might be a possible explanation for the unexpected threshold behaviour of l-changing collision times inferred in \rref{schlagmueller:ucoldchemreact:prx}.

While presently intractable, a more complete calculation should include all perturbers in the orbital volume. This likely provides additional electronic final states, including superpositions of trilobite orbitals \cite{Eiles_highdensenv_2016}. However since we find numerically that the volume covered by the Rydberg electron in \fref{trilobite_slowdown}{b} almost certainly contains dense clusters of a few condensate atoms, it appears likely based on \cite{Luukko_polytrilob_PRL} that the electron will relocate to one of these clusters. 


\end{document}